\documentclass[prd, twocolumn, showpacs, superscriptaddress, floatfix, nofootinbib]{revtex4}

\usepackage[dvips]{graphicx}
\usepackage{graphicx}
\usepackage{epsf}
\usepackage{amsmath}
\usepackage{amssymb}

\usepackage{graphicx}
\usepackage{dcolumn}
\usepackage{bm}
\def\be{\begin{equation}}
\def\ee{\end{equation}}
\def\bea{\begin{eqnarray}}
\def\eea{\end{eqnarray}}

\usepackage{color}
\newcommand{\tc}{\textcolor{black}}

\begin{document}

\title{Higgs Boson in RG running Inflationary Cosmology}

\author{Yi-Fu Cai}
\email{ycai21@asu.edu}
\affiliation{Department of Physics, Arizona State University, Tempe, AZ 85287}
\affiliation{Department of Physics, McGill University, Montr\'eal, QC, H3A 2T8, Canada}
\author{Damien A. Easson}
\email{easson@asu.edu}
\affiliation{Department of Physics, Arizona State University, Tempe, AZ 85287}


\begin{abstract}
An intriguing hypothesis is that gravity may be non-perturbatively renormalizable via the notion of asymptotic safety. We show that the Higgs sector of the SM minimally coupled to asymptotically safe gravity can generate the observed near scale-invariant spectrum of the Cosmic Microwave Background through the curvaton mechanism. The resulting primordial power spectrum places an upper bound on the Higgs mass, which for {\tc{finely tuned}} values of the curvaton parameters, is compatible with the recently released Large Hadron Collider data.
\end{abstract}

\maketitle

\newcommand{\eq}[2]{\begin{equation}\label{#1}{#2}\end{equation}}

\section{Introduction}

Weinberg has suggested that the effective description of a quantum gravitational theory may be nonperturbatively renormalizable through the notion of asymptotic safety (AS) \cite{Weinberg:1977, Weinberg:1979}. In such a scenario the renormalization group (RG) flows approach a fixed point in the ultraviolet (UV) limit, and a finite dimensional critical surface of trajectories evolves to this point at short distance scales \cite{Weinberg:2009bg, Weinberg:2009wa}. Such a fixed point was found in the Einstein-Hilbert truncation \cite{Reuter:1996cp}, and the scenario was studied extensively in the literature \cite{Souma:1999at, Bonanno:2001xi, Lauscher:2001ya, Litim:2003vp, Codello:2006in, Cai:2010zh} (for recent reviews see \cite{Niedermaier:2006ns, Percacci:2007sz}).

Inflationary cosmology is the most promising candidate theory for describing the early universe \cite{Guth:1980zm, Linde:1981mu, Albrecht:1982wi}. The paradigm solves the homogeneity, flatness, horizon and unwanted relic problems. It also predicts a nearly scale-invariant primordial power spectrum, in agreement with the data from modern cosmological experiments \cite{Komatsu:2010fb}. However, the model generally requires an
as of yet unobserved scalar field. The Higgs boson is a scalar field predicted by the SM of particle physics and exciting evidence for its existence has recently been released from the Large Hadron Collider (LHC) experiment \cite{LHC:2011, ATLAS:2011, CMS:2011}. In the past, it was hoped that the Higgs field might play a dual role as the inflaton, but the corresponding energy scale is much lower than what inflationary cosmology typically requires. Consequently, a model of Higgs inflation was proposed in which the Higgs field is non-minimally coupled to Einstein gravity \cite{Bezrukov:2007ep, Bezrukov:2008ej, Bezrukov:2009db, Barvinsky:2008ia, Barvinsky:2009fy, Barvinsky:2009ii, DeSimone:2008ei} (see earlier attempt in \cite{CervantesCota:1995tz}).
It was soon realized that such non-minimal coupling leads to an energy scale for unitarity violation which is expected to be larger than the inflationary scale; otherwise, the effective field description would fail (for example, see \cite{Burgess:2009ea, Barbon:2009ya, Lerner:2009na, Burgess:2010zq, Bezrukov:2010jz, Atkins:2010yg, Horvat:2011wr} for detailed discussions). Therefore, difficult challenges exist for models which attempt to use the Higgs field to drive inflation.

In this paper, we propose that the Higgs boson may play an important role in the early inflationary universe if the gravitational theory is asymptotically safe. {\tc{Although at present there is no explicit proof of the asymptotical safety, it deserves to ask whether the gravity theories under asymptotical safety are testable in any experiment. The current study is exactly based on a specific truncation of the gravity action and certain form of RG equations according to the conjecture of asymptotical safety.}} In the frame of AS gravity, the gravitational constant $G$ and cosmological constant $\Lambda$ are running along with the energy scale, and thus vary throughout the cosmological evolution. It has been argued that if there are no intermediate energy scales between the SM and AS scales, the mass of the Higgs boson is predicted to be $m_H=126 \,{\rm GeV}$ with only several {\rm GeV} uncertainty \cite{Shaposhnikov:2009pv}. We find a suitable inflationary solution can be obtained in a cosmological system which contains a Higgs boson and AS gravity, along the lines of \cite{Cai:2011kd}. Different from the idea of \cite{Atkins:2010yg}, in our model the inflationary solution can be obtained purely in the frame of AS gravity without the help of the Higgs boson, but the Higgs field plays the role of seeding primordial power spectrum which explains the CMB observation. In this model, the gravitational constant and cosmological constant vary along with the running of the cutoff scale and thus contribute extra term to the energy stress tensor. To take into account of the extra term, the model is found to be corresponding to a $f(R)$ model \cite{Cai:2011kd} (see also \cite{Bonanno:2012jy, Hindmarsh:2012rc} for generalized discussions). As a consequence, there are effectively two scalar degrees of freedom, one being the adiabatic mode and the other being an iso-curvature mode.

We find the corresponding perturbation theory leads to both the primordial power spectrum for the curvature perturbation and the entropy perturbation. When the cutoff scale runs lower than a critical value, inflation abruptly ends and the Higgs field can give rise to a reheating phase. During this phase, the fluctuations seeded by the Higgs field can be converted into the curvature perturbation through the curvaton mechanism \cite{Lyth:2001nq} (see also \cite{Mollerach:1989hu, Linde:1996gt, Enqvist:2001zp}). We derive a relation between the spectral index of the primordial power spectrum and the Higgs mass. We confront this relation with the latest cosmological observations and collider experiment data, and find they are consistent under a group of canonical values of curvaton parameters.

In this paper, we will work with the reduced Planck mass, $M_{p} = 1/\sqrt{8\pi G_N}$, where $G_N$ is the gravitational constant in the IR limit, and adopt the mostly-plus metric sign convention $(-,+,+,+)$.

\section{The model}

Consider the SM of particle physics minimally coupled to gravity
\begin{eqnarray}\label{S_J}
 S = \int d^4x\sqrt{-g} \bigg[ \frac{R-2\Lambda}{16\pi G} + {\cal L}^{SM} \bigg]~.
\end{eqnarray}
In this AS gravity frame, the gravitational constant $G$ and the cosmological constant $\Lambda$ vary along the cutoff scale $p$. The running behaviors are approximately described by
\begin{eqnarray}\label{GL_run}
 G(p)^{-1} &\simeq& G_{N}^{-1}+\xi_G p^2~,\\
 \Lambda(p) &\simeq& \Lambda_{IR} + \xi_\Lambda p^2~,
\end{eqnarray}
where $G_{N}$ and $\Lambda_{IR}$ are the values of gravitational constant and cosmological constant in the IR limit. The coefficients $\xi_G$ and $\xi_\Lambda$ are determined by the physics near the UV fixed point of RG flows in AS gravity.

The scalar sector of the SM contains the Higgs boson. We use the unitary gauge for the Higgs boson $H = \frac{h}{\sqrt{2}}$ and neglect all gauge interactions for the time being. In this case, the Lagrangian of the Higgs field is given by,
\begin{eqnarray}\label{L_H}
 {\cal L}^{SM} \supseteq - \frac{1}{2}\partial_\mu h\partial^\mu h -V(h)~,
\end{eqnarray}
where $V(h)$ is the potential of the Higgs field, which is typically in form of $\frac{\lambda}{4}(h^2-v^2)^2$.

Varying the Lagrangian with respect to the metric, one derives the generalized Einstein equation,
\begin{eqnarray}\label{EoM_J}
 R_{\mu\nu}-\frac{R}{2}g_{\mu\nu}+\Lambda g_{\mu\nu} = 8\pi G (T_{\mu\nu}^{SM} + T_{\mu\nu}^{AS})~.
\end{eqnarray}
Here the RG running of $G$ can effectively contribute to the stress energy tensor through
\begin{eqnarray}
 T_{\mu\nu}^{AS} = (\nabla_\mu\nabla_\nu-g_{\mu\nu}\Box)(8\pi G)^{-1}~,
\end{eqnarray}
where we have introduced the covariant derivative $\nabla_\mu$ and the operator $\Box \equiv -g^{\mu\nu} \nabla_\mu\nabla_\nu$. The Higgs field $h$ obeys the Klein-Gordon equation. Additionally, the running of cutoff scale is controlled by the Bianchi identity, which requires,
\begin{eqnarray}
 (R-2\Lambda)\frac{\nabla_\mu G}{G} + 2\nabla_\mu\Lambda = 0~.
\end{eqnarray}
Consequently, the dynamics of this cosmological system are completely determined. Explicitly, from the analysis of Ref. \cite{Cai:2011kd}, the Ricci scalar and the cutoff scale are able to be identified as
\begin{eqnarray}
 R = 2(\Lambda_{IR}-\frac{\xi_\Lambda}{\xi_G G_N})+\frac{4G\xi_\Lambda}{\xi_G}~.
\end{eqnarray}
Then we can reformulate the original theory as an $f(R)$ model as follows,
\begin{eqnarray}
 S = \int d^4x\sqrt{-g} \bigg[ f(R) + {\cal L}^{SM} \bigg]~,
\end{eqnarray}
with
\begin{eqnarray}
 f(R) = \frac{[G_N(R-2\Lambda_{IR})+2Z]^2}{128\pi G_N^2 Z}~,
\end{eqnarray}
where we have introduced $Z={\xi_\Lambda}/{\xi_G}$.

\section{Inflationary Cosmology}

\subsection{Background Dynamics}

We now turn our attention to early universe inflationary solutions. This system is most easily studied by making a conformal transformation,
\begin{eqnarray}\label{g_E}
 \tilde{g}_{\mu\nu} = \Omega^2 g_{\mu\nu}~,~~\Omega^2=\frac{G_N}{G}~,
\end{eqnarray}
where $\Omega^2$ is the conformal factor with a new scalar field
\begin{eqnarray}
 \phi\equiv -\frac{\sqrt{6}M_p}{2} \ln\frac{G_N}{G}~,
\end{eqnarray}
being introduced.

The original system, therefore, is equivalently described in terms of two scalar fields minimally coupled to Einstein gravity without RG running,
\begin{eqnarray}\label{L_E}
 {\cal L} = \frac{\tilde{R}}{16\pi G_N} - \frac{1}{2}(\tilde\nabla \phi)^2 - \frac{e^{2b(\phi)}}{2}(\tilde\nabla h)^2 -\tilde{V}(\phi, h)~,
\end{eqnarray}
where the factor $b(\phi)\equiv\frac{\phi}{\sqrt{6}M_p}$ and $\tilde{V}(\phi, h) = U(\phi)+e^{4b}V(h)$, with
\begin{eqnarray}
 U(\phi) \simeq 8\pi M_p^4 \bigg[ \frac{\xi_\Lambda}{\xi_G} \bigg( 1-e^{2b(\phi)} \bigg) + G_N\Lambda_{IR} \, e^{2b(\phi)} \bigg]~.
\end{eqnarray}
Note that, the form of $\phi$'s potential is derived from the RG running of the AS gravity \cite{Cai:2011kd}. The last term of $U(\phi)$ is proportional to $G_N\Lambda_{IR}$. Substituting the observed values of $G_N$ and $\Lambda_{IR}$, we find this term is insignificant throughout the past cosmological evolution. Hence, we make the approximation
\begin{eqnarray}\label{U_potential}
 U(\phi)\simeq 8\pi M_p^4 \frac{\xi_\Lambda}{\xi_G} \left[1-e^{2b(\phi)} \right]~,
\end{eqnarray}
which is a sufficiently flat inflationary potential in the regime where $b(\phi)\ll-1$.

We denote the frame proceeding the conformal transformation as the Einstein frame (despite the non-canonical form of the $h$ kinetic term). Substitution of the flat Friedmann-Robertson-Walker (FRW) metric, $ds^2=-dt^2+a^2(t)d\vec{x}^2$, leads to the Friedmann equations:
\begin{eqnarray}
 H^2 &=& \frac{1}{3M_p^2} (\frac{\dot\phi^2}{2}+\frac{e^{2b}\dot h^2}{2}+\tilde{V} )~~ {\rm and} \\
 \dot{H} &=& -\frac{1}{2M_p^2} (\dot\phi^2+e^{2b}\dot h^2 )~,
\end{eqnarray}
where we have defined the Hubble parameter $H\equiv\frac{\dot a}{a}$ and the dot denotes the time derivative in the Einstein frame. The coupled Klein-Gordon equations for the two scalars are:
\begin{eqnarray}
 && \ddot\phi + 3H\dot\phi + \tilde{V}_{,\phi} = b_{,\phi}e^{2b}\dot{h}^2~,\\
 && \ddot{h} + (3H+2b_{,\phi}\dot\phi)\dot{h} + e^{-2b}\tilde{V}_{,h} = 0~.
\end{eqnarray}

\subsection{Inflationary Solution}

To search for a successful inflationary solution, we introduce a series of slow roll parameters,
\begin{eqnarray}\label{slow roll}
 \epsilon_\phi = \frac{\dot\phi^2}{2M_p^2H^2}~,~\epsilon_h = \frac{e^{2b}\dot{h}^2}{2M_p^2H^2}~,~\eta_{IJ} = \frac{\tilde{V}_{,IJ}}{3H^2}~.
\end{eqnarray}
The subscript ``$_{,I}$" denotes the derivative with respect to the $I$th--field  (with $I$ being $\phi$ or $h$). During inflation, these parameters are required to be less than unity. However, the key parameters need to yield a successful inflationary background are associated with the scalar $\phi$, i.e., $\epsilon_\phi$ and $\eta_{\phi\phi}$. This is because the potential for $\phi$ is flat in the regime $\phi\ll-M_p$ and correspondingly the parameters related to $h$ are suppressed by the small-valued factor $e^{2b}$.

The background dynamics is determined by the following solutions (under slow roll approximation),
\begin{eqnarray}\label{solution_sl}
 \dot\phi \simeq -\frac{U_{,\phi}}{3H}~,~\dot{h} \simeq -\frac{e^{2b}V_{,h}}{3H}~,~H^2\simeq\frac{U}{3M_p^2}~.
\end{eqnarray}
Inflation ends when $\epsilon_\phi = 1$. Combing this condition with the background solution for $\dot\phi$ in (\ref{solution_sl}), we find the value of $\phi$ at the end of inflation: $\phi_f \simeq -0.56 M_p$. The number of e-folding of inflation is given by ${\cal N}=-\int_i^f\frac{Ud\phi}{M_p^2U_{,\phi}}$, so that
\begin{eqnarray}
 {\cal N}(\phi) \simeq \frac{3}{2} e^{-2b(\phi)}+3b(\phi)-1.68~.
\end{eqnarray}
To obtain ${\cal N}=60$, we require the initial inflaton value to be, $\phi_i \simeq -4.7M_p$, which lies in the regime where RG flows of AS gravity have approached the UV fixed point.

\subsection{Graceful exit}

Eventually, the slow roll conditions are violated, when $\phi$ reaches $\phi_f$. Consequently, $\phi$  enters a period of fast roll, and finally approaches $\phi=0$ at which point the AS gravity reduces to traditional Einstein gravity. We suggest two possible reheating mechanism. One is that the inflaton $\phi$ decays to radiation, driving the universe to a phase of thermal expansion directly; the other possibility is that the universe reheats only after the energy scale drops sufficiently so that the SM Higgs boson is responsible for the reheating process. {\tc{Note that, in the latter case, after inflation the inflaton would experience a period of fast roll during which its effective equation of state approximately equals to $1$. However, the energy density of the Higgs field scales as radiation due to the $\lambda h^4$ potential. As a consequence, the relative fraction of Higgs boson could grow after inflation even in the case that the Higgs has large coupling to other standard model particles. }}

{\tc{The decay of the inflaton $\phi$ into radiation strongly depends on the interaction between them. The detailed reheating process could be realized by the first mechanism or the second, or even the combination of these two. However, in the present paper we will not address on the details but simply consider the occupation of the Higgs boson at the last reheating surface as a free parameter.}}
Assuming that the Higgs boson would decay into radiation instantly at the last reheating surface {\tc{(regardless which mechanism)}}, the reheating temperature is approximately given by $T_{re} \simeq (\frac{2\lambda}{\pi^2g_d})^{\frac{1}{4}} h_{re}$ where $h_{re}$ is the value of $h$ at the reheating surface, and $g_d\simeq106.75$ is the number of degrees of freedom of the SM. After inflation but before reheating, $h$  oscillates along the potential $V\sim \lambda h^4$. Thus the value of $h$ can be restricted to $v<|h_{re}|<h_*$ with $h_*$ being the value of the Higgs boson at the moment of Hubble-exit.

\section{Higgs Curvaton}

\subsection{Field fluctuations}

During inflation, the background dynamics are not affected by the Higgs field, however its quantum fluctuations are able to source a nearly scale-invariant entropy perturbation. In a two field inflationary model, we decompose the field variables into a background part and fluctuations: $\phi\rightarrow\phi+\delta\phi$ and $h\rightarrow h+\delta{h}$. The field fluctuations combine to give adiabatic and iso-curvature modes as:
\begin{eqnarray}
 \delta\sigma &=& \cos\theta\delta\phi + \sin\theta e^b\delta{h}~,\\
 \delta{s} &=& -\sin\theta\delta\phi +\cos\theta e^b\delta{h}~,
\end{eqnarray}
with the trajectory angle being defined by $\cos\theta = \frac{\dot\phi}{\dot\sigma}$ and $\sin\theta = \frac{e^b\dot{h}}{\dot\sigma}$ \cite{DiMarco:2002eb}. To take into account the metric fluctuation (the gravitational potential $\Phi$) we introduce the canonical perturbation variables,
\begin{eqnarray}
 v_\sigma=a(\delta\sigma+\frac{\dot\sigma}{H}\Phi)~,~~
 v_s=a\delta s~,
\end{eqnarray}
which characterize gauge-invariant adiabatic and iso-curvature perturbations. Up to leading order in the slow roll approximations, these two variables obey the perturbation equations \cite{Mukhanov:1990me}:
\begin{eqnarray}
 v_{\sigma(s)}''+(k^2-\frac{a''}{a})v_{\sigma(s)} \simeq 0~,
\end{eqnarray}
where the prime denotes differentiation with respect to conformal time, $\tau\equiv\int\frac{dt}{a}$. Solving this equation in the inflationary phase, we can obtain nearly scale-invariant primordial power spectra for adiabatic and iso-curvature perturbations, and their corresponding amplitudes are $|\delta\sigma|\simeq|\delta s|\simeq\frac{H_*}{2\pi}$ at the Hubble-exit moment $t_*$. Hence, the amplitudes of the field fluctuations at this moment are,
\begin{eqnarray}\label{delta h}
 |\delta\phi_*|\simeq \frac{H_*}{2\pi}~,~~|\delta{h}_*|\simeq \frac{H_*}{2\pi e^b_*}~.
\end{eqnarray}

\subsection{The curvaton mechanism}

When inflation ends, $\phi$ rapidly approaches the IR limit of the AS gravity. Because $\phi$ generally couples to other matter fields through the conformal factor $\Omega^2$, we expect radiation to be produced following the inflationary phase. Recall that, the Higgs boson can survive during inflation due to the slow roll conditions and then starts to oscillate along the $\lambda h^4$ potential. Consequently, the universe is dominated by both the radiation and the Higgs boson after inflation\footnote{\tc{To be precise, there exists a short period during which the universe is dominated by a fast-roll $\phi$ with $w_\phi\simeq1$. However, as the contribution of $\phi$ would dilute out soon, only radiation and the Higgs boson could be left after the fast-roll phase. }}. This process is analogous to the familiar curvaton scenario. Instead of a matter-like curvaton oscillation with $w=0$ as studied in \cite{Lyth:2001nq, Lyth:2002my}, a $\lambda h^4$ potential yields an effective equation of state for the Higgs boson, which is the same as radiation with $w_h=\frac{1}{3}$ \cite{Turner:1983he, Huang:2008zj}. Thus, this generalized curvaton mechanism \cite{Cai:2010rt} may be used to generate the primordial curvature perturbation in agreement with the current CMB measurements.

We begin by writing down the relation between the curvaton fluctuation $\delta{h}$ and its curvature perturbation $\zeta_{h}$. Choosing the spatially flat slice for the Higgs curvaton, one finds $\rho_h=\bar\rho_h e^{3(1+w_h)\zeta_h}$ in the neighborhood of the curvaton reheating hypersurface. Consider the curvaton perturbation generated from vacuum fluctuations inside the Hubble radius. These fluctuations satisfy a Gaussian distribution at the Hubble-exit. In general, the Hubble-exit value of the Higgs boson $h_*$ can be related to the initial amplitude of curvaton oscillation $h_o$ through a model-dependent function $h_o=g(h_*)$. For example, in the present model, if the curvaton starts to oscillate immediately after inflation, $h_o\simeq h_*$; however, if there is a short slow rolling behavior for $h$ following inflation,  $h_o \simeq \frac{11}{12}h_*$. In this case, the curvature perturbation of the Higgs field in the oscillating phase is given by,
\begin{eqnarray}\label{zeta_h}
 \zeta_h = \frac{\delta\rho_h}{3(1+w_h)\rho_h} \simeq q_h \frac{\delta h_*}{h_*}~,
\end{eqnarray}
with $q_h\equiv\frac{h_oh_*}{h_o^2-v^2}$. The coefficient $q_h$ can be further simplified as $q_h\simeq \frac{h_*}{h_o}$ when $|h_o|\gg v$ if curvaton reheating occurs at energy scales higher than the SM scale.

We now need to relate $\zeta_h$ to $\zeta$. In the sudden decay approximation, the relation can be computed analytically. Consider the case that the Higgs boson decays on a uniform total density hypersurface. On this slice we have $\rho_h+\rho_r=\rho_T$ where $\rho_r$ and $\rho_T$ denote the energy density of radiation and that of the total system, respectively. Making use of the expression for the curvature perturbation on a uniform density slice, we find $\rho_r=\bar\rho_r e^{4(\zeta_r-\zeta)}$ and $\rho_h=\bar\rho_h e^{3(1+w_h)(\zeta_h-\zeta)}$ during curvaton oscillation. Thus, $\zeta$ and $\zeta_h$ are related on the reheating hypersurface as follows,
\begin{eqnarray}
 (1-\Omega_h)e^{4(\zeta_r-\zeta)} + \Omega_h e^{3(1+w_h)(\zeta_h-\zeta)} =1~,
\end{eqnarray}
where $\Omega_h=\rho_h/\rho_T$ is the dimensionless density parameter for the curvaton. For the curvaton mechanics to succeed, we must assume that the fluctuation $\zeta_r$ seeded by the inflaton field is negligible. We will address this concern below and now turn our attention to the Higgs curvaton. Therefore, we have
\begin{eqnarray}\label{zeta}
 \zeta = q_T \zeta_h~,~~q_T=\frac{3(1+w_h)\Omega_h}{4-(1-3w_h)\Omega_h}~,
\end{eqnarray}
and in our explicit case, $q_T=\Omega_h$ at the curvaton decay surface.

Combining Eqs. (\ref{zeta_h}) and (\ref{zeta}) and the field fluctuation (\ref{delta h}), we obtain the primordial power spectrum of curvature perturbation seeded by the Higgs boson,
\begin{eqnarray}\label{P_zeta}
 P_\zeta = \frac{q_h^2q_T^2}{4\pi^2e^{2b_*}}\frac{H_*^2}{h_*^2}~.
\end{eqnarray}
We see from  (\ref{P_zeta}), that the final curvature perturbation depends on five parameters: $q_h$, $q_T$, $H_*$, $h_*$, and $e^{b_*}$. Compared with the usual curvaton mechanism, our model contains a new parameter $e^{b_*}$ due to the conformal transformation made in Eq. (\ref{g_E}). However, since the background dynamics of inflation are driven by the RG running of AS gravity, we find $e^{b_*}\simeq 0.15$ for observable perturbation modes at Hubble-exit. Moreover, we have specified the curvaton to be the Higgs boson and hence, the potential is of an explicit form, and thus $q_h\sim 1$. Since the latest CMB data reveals $P_\zeta\simeq2.4\times10^{-9}$, we can deduce the useful relation $ H_* \simeq 4.5\times10^{-5} \frac{h_*}{q_T}$. Subsequently, we are left to constrain $q_T$, $H_*$ and $h_*$ by various theoretical and observational requirements. In the following, we calculate the tensor-to-scalar ratio, the spectral index, and the reheating temperature, respectively, and then constrain the remaining parameters.

\section{Constraints}

The calculation of primordial tensor perturbations is identical to that of ordinary inflationary models, and thus the tensor power spectrum is nearly scale-invariant,  given by $P_t = \frac{2H_*^2}{\pi^2M_p^2}$. The tensor-to-scalar ratio:
\begin{eqnarray}
 r \equiv \frac{P_t}{P_\zeta}=\frac{8e^{2b_*}h_*^2}{q_h^2q_T^2M_p^2}~.
\end{eqnarray}
Since $e^{b_*}\simeq0.15$ and $q_h\sim 1$, we obtain $r\simeq\frac{0.2h_*^2}{q_T^2M_p^2}$. According to the latest CMB data, $r$ is required to be less than $0.36$ \cite{Komatsu:2010fb}, so that $h_*<2q_TM_p$.
A further constraint comes from the curvaton condition that the contribution of inflaton fluctuation to curvature perturbation should be negligible. This condition requires $q_T\zeta_h\gg(1-q_T)\zeta_r$. Since $\zeta_r$ corresponds to the radiation perturbation,  inherited from inflaton fluctuation, $\zeta_r\simeq \frac{H_*}{2\pi\sqrt{2\epsilon_*}M_p}$. Choosing a group of canonical parameter values, this condition requires $h_*<q_TM_p$ which is close to the observational bound provided by the tensor-to-scalar ratio.  Combining this inequality and the expression of $H_*$ derived from $P_\zeta$, one obtains a constraint on the inflationary Hubble parameter $H_*<4.5\times10^{-5}M_p$.

During inflation the Hubble parameter is approximately constant and the field fluctuations are approximately conserved after Hubble exit. The spectral tilt $n_h \equiv 1+\frac{d\ln P_{h}}{d\ln k}$ of the primordial perturbations at the moment of horizon crossing is:
\begin{eqnarray}\label{n_h}
 n_h \simeq 1 -\frac{2\epsilon_*^{\frac{1}{2}}}{\sqrt{3}} -2\epsilon_* +\frac{q_h^2q_T^2m_H^2}{4\pi^2v^2P_\zeta}~,
\end{eqnarray}
where in the r.h.s. we have used the expression for the primordial power spectrum. The Higgs mass is determined by $m_H=\sqrt{2\lambda}v$. In this model the inflaton potential is explicitly defined, and yields $\epsilon\simeq1.6\times 10^{-4}$ at the beginning of inflation. Thus, the spectral index can be simplified as $n_h\simeq 0.985+\frac{q_h^2q_T^2m_H^2}{4\pi^2v^2P_\zeta}$ for perturbation modes which exit the Hubble radius during the first several efolds. As a consequence, the Higgs mass acquires an upper bound from the CMB measurement:
\begin{eqnarray}
 m_H<3\times10^{-5}\frac{v}{q_T}~.
\end{eqnarray}
For the SM Higgs, $v\simeq246{\rm GeV}$, and if $q_T$ is smaller than $5.7\times 10^{-5}$,  we conclude that the Higgs mass has to be less than $129 {\rm GeV}$.

\section{Conclusions and Discussion}

In this paper, we proposed a new inflation model in the context of an asymptotically safe gravitational cosmology with renormalization group running gravitational constant. The theory is minimally coupled to the Standard Model sector. As observed in \cite{Shaposhnikov:2009pv}, if there are no intermediate energy scales between the SM and the AS scales, the mass of the Higgs boson is determined by a fixed point and is approximately $126 {\rm GeV}$. This prediction is coincident with the recent results announced from the ATLAS and CMS experiments, indicating that the Higgs mass is in the range $116 - 131 {\rm GeV}$ (ATLAS\cite{ATLAS:2011}) or $115 - 127 {\rm GeV}$ (CMS\cite{CMS:2011}), with other masses excluded at the $95\%$ confidence level.
%
%
{\tc{In the frame of the AS inflation model, we introduce the Higgs boson to play the role of a curvaton which is responsible for generating the primordial curvature perturbation.}}
%
%
Furthermore, if the occupation of curvaton density at the reheating surface is fixed, then observational constraints on the spectral index imply an upper bound on the Higgs mass. After a fine tuning of the curvaton parameters, the model is consistent with recent LHC data. A complete data fitting of our model will appear in the forthcoming work \cite{ycde2012b}.

We conclude with a few remarks concerning the details of the model as follows.

The first involves the estimate of the values of $\xi_\Lambda$ and $\xi_G$ from the AS gravity. In a general case, AS gravity implies this parameter is typically less than unity but strongly depends on the detailed parametrization of RG flows. Namely, in the simplest parametrization in which the UV and IR behaviors are connected in a linear approximation, the ratio of $\xi_\Lambda$ and $\xi_G$ often leads to inflationary scale which is much higher than the GUT scale, and thus the inflation model based on this example suffers from a fine tuning problem~\cite{Cai:2011kd}. However, one may consider RG flows which are more general than the linear approximation. For example, one can parameterize the RG flows by a log approximation as suggested in Ref. \cite{Bonanno:2012jy}. In this case, the parameter space is significantly larger and a very small value of $\xi_\Lambda / \xi_G$ is allowed.

Our second remark concerns the estimate of the values of the fields $\phi$ and $h$ at the beginning of inflation. In principle, they may be completely determined by numerical calculations. We may choose the inflaton $\phi$ to initially be about a few times the Planck mass
in order to attain a sufficiently long inflationary period. Interestingly, these values are also consistent with observational constraints given in the present paper\footnote{\tc{Note that the scenario of Higgs curvaton is recently analyzed in \cite{Choi:2012cp} and some stringent bound on primordial non-Gaussianity is achieved.}}.

{\tc{The third is on the relation between the spectral index of the primordial power spectrum and the Higgs mass obtained in Eq. \eqref{n_h}. We should be aware of that this relation is only derived at tree level in the present paper. However, one would expect the relation to be altered when radiative corrections are taken into account. This is an interesting topic which we would like to address in the following-up project.
}}

\begin{acknowledgments}
We are grateful to R. Brandenberger for discussions. We also thank the anonymous referee for valuable comments and introducing relevant references on the non-minimally coupled Higgs inflation model and \cite{Choi:2012cp}. The work of Y.F.C. and D.A.E is supported in part by the DOE and the Cosmology Initiative at Arizona State University.
\end{acknowledgments}

\end{document}